\documentclass[utf8]{FrontiersinHarvard} 

\usepackage{url,hyperref,lineno,microtype,subcaption}
\usepackage[onehalfspacing]{setspace}
\usepackage{siunitx}
\usepackage{booktabs}
\usepackage{soul}
\setstcolor{red}

\graphicspath{{figures/}}

\def\keyFont{\fontsize{8}{11}\helveticabold }
\def\firstAuthorLast{B K Jha {et~al.}} 
\def\Authors{Bibhuti~Kumar~Jha\,$^{1,2,3,*}$, Manjunath~Hegde\,$^{1,2}$, Aditya~Priyadarshi\,$^{1, 2}$, Sudip~Mandal\,$^{4}$, B~Ravindra\,$^{1}$ and Dipankar~Banerjee\,$^{1,2,5,*}$}
\def\Address{$^{1}$Aryabhatta Research Institute of Observational Sciences, Nainital, Uttarakhand, India\\
$^{2}$Indian Institute of Astrophysics, Bangalore, Karnataka, India\\
$^{3}$Pondicherry University, Kalapet, Puducherry, India\\
$^{4}$Max Planck Institute for Solar System Research, Göttingen, Germany\\
$^{5}$Center of Excellence in Space Sciences India, IISER Kolkata, Mohanpur, West Bengal, India}
\def\corrAuthor{Bibhuti Kumar Jha}

\def\corrEmail{bibhuti@aries.res.in; maitraibibhu@gmail.com}

\newcommand{\aap}{    {\it Astron. Astrophys.}}

\newcommand{\aapr}{   {\it Astron. Astrophys. Rev.}}

\newcommand{\apj}{    {\it Astrophys. J.}}

\newcommand{\mnras}{  {\it Mon. Not. Roy. Astron. Soc.}}

\newcommand{\solphys}{{\it Solar Phys.}}
 
\newcommand{\ssr}{    {\it Space Sci. Rev.}}

\newcommand{\lrsp}{  {\it Living Rev. Sol. Phys.}}

\begin{document}
\onecolumn
\firstpage{1}
\title[Extended Sunspot Area Series from KoSO]{Extending the Sunspot Area Series from Kodaikanal Solar Observatory} 

\author[\firstAuthorLast ]{\Authors} 
\address{} 
\correspondance{} 

\extraAuth{Dipankar Banerjee \\ dipu@aries.res.in}

\maketitle

\begin{abstract}


Kodaikanal Solar Observatory (KoSO) possesses one of world's longest and homogeneous records of sunspot observations that span more than a century (1904\,--\,2017). Interestingly, these observations (originally recorded in photographic plates/films) were taken with the same setup over this entire time period which makes this data unique and best suitable for long-term solar variability studies. A large part of this data, between 1921\,--\,2011, were digitized earlier and a catalog containing the detected sunspot parameters (e.g., area and location) was published in \citet{Mandal2017a}. In this article, we  extend the earlier catalog by including new sets of data between 1904\,--\,1921 and 2011\,--\,2017. To this end, we digitize and calibrate these new datasets which include resolving the issue of random image orientation. We fix this by comparing the KoSO images with  co-temporal data from Royal Greenwich Observatory. Following that, a semi-automated sunspot detection and automated umbra detection algorithm are implemented onto these calibrated images to detect sunspots and umbra. Additionally, during this catalog updation, we also filled data gaps in the existing KoSO sunspot catalog (1921\,--\,2011) by virtue of re-calibrating the ‘rouge’ plates. This updated sunspot area series covering nearly 115 years (1904\,--\,2017) are being made available to the community and will be a unique source to study the long term variability of the Sun.

\tiny
 \keyFont{ \section{Keywords:} Sun, Sunspots, Solar Cycle, Kodaikanal Solar Observatory, white-light, sunspot area, umbra area} 
\end{abstract}

\section{Introduction}
Sunspots have always been a central part of our understanding of the Sun and its long-term variability \citep{Solanki2003}. The systematic and methodical observation of these spots has revealed that their appearance is periodic, with a periodicity of around 11 years, known as the solar cycle or solar activity cycle \citep{Schwabe1844, Hathaway2015}. Furthermore, the magnetic nature of these spots \citep{Hale1908, Parker1955a} makes them an ideal proxy for understanding solar magnetism and its complex variability \citep{Tlatov2014,Nagovitsyn2017}. Today, it is well established that the solar activity cycle is governed by the solar dynamo process operating in the convection zone of the Sun \citep{Parker1955, Parker1975, Charbonneau2010}. In the dynamo process, the solar activity cycle is the manifestation of the periodic nature of the large-scale solar magnetic field (poloidal $\rightleftharpoons$ toroidal). As a consequence, the number of sunspots and the corresponding area covered by these spots on the solar surface is dictated by the strength of the toroidal field generated in the dynamo process. Hence, the historical observation of these spots carries a vital information about the nature of toroidal field in the past and it will be crucial for the reconstruction of historical global solar magnetic field \citep{Jiang2011,Jiang2014a}. Apart from that, sunspots and solar activity are also intrinsically linked to solar transient events such as solar flares and coronal mass ejections (CMEs) and their frequency of occurrence. Since these transients are the primary drivers for the space weather condition, the historical observation will have crucial role for the understanding of space weather condition in past.

Over the last 150 years, many observatories around the world have begun regular observations of the Sun. The Royal Observatory of Greenwich \citep[RGO;][]{Willis2013} had been the leader in such a campaign having the record of white-light observations from 1874 to 1976, which was later continued using the Solar Optical Observing Network (SOON) by the US Air Force (USAF). After a few decades, in 1904, Kodaikanal Solar Observatory \citep[KoSO;][]{Hasan2010} also joined this campaign and started regular observation of the Sun on photographic plates/films in multi-wavelengths (white-light since 1904, Ca-K since 1904 and H-$\alpha$ since 1912), independently in India. In particular, the white-light observation taken at KoSO provides one of the most homogeneous data series for over 100~years. Since these observations are taken from the same location and using the same telescope (since 1918) for such an extended period, KoSO provides a unique data series ideal for the long-term study of the Sun. Although the white-light data had been digitized \citep{Ravindra2013} since 1904; owing to calibration issues, it has not been utilized for the period 1904\,--\,1920. Therefore, the area series \citep{Mandal2017a}, the study of penumbra to umbra area ratio \citep[1923\,--\,2011;][]{Jha2018, Jha2019} and solar differential rotation \citep[1923\,--\,2011;][]{Jha2021} were limited to the period 1921\,--\,2011.

In-spite of the fact that RGO is the only observatory in the world with white-light data for 1904\,--\,1920 (Cycle-14 and Cycle-15), these images are available either in low resolution or in the form of drawings\footnote{\url{http://fenyi.solarobs.csfk.mta.hu/GPR/index.html}}. In this regard, the high-resolution white-light data available from KoSO serves as the only data set for this time frame. It is almost impossible for the ground based observatories to have a uniform and homogeneous data because of the varying atmospheric conditions and bad weather. Hence,  in the last 30~years, there has been considerable effort to make the homogeneous sunspot area series by cross-correlating the sunspot area from various observatories, e.g., \citet{Fligge1997}, \citet{Baranyi2001, Baranyi2013}, \citet{Balmaceda2009} and recently \citet{Mandal2020}. In all these studies, RGO data has been used primarily as a reference to cross-correlate other data because of its outstanding data coverage. Now, the availability of KoSO data will complement the existing data series and will be helpful in cross-correlating the RGO and KoSO data in these initial overlapping periods. In this article, we present the extension of the KoSO sunspot area series reported in \citet{Mandal2017a} for 1904\,--\,2017 along with the umbral area after resolving the calibration issues for the initial~17 years of data. We have also included data for the period of 2012\,--\,2017, which was previously not reported due to ongoing digitization. In this article we present the updated data statistics, the issues with the calibration and the resolution of these issues in Section~\ref{s:data}. In Section~\ref{s:result}, we will discuss the updated area series and its comparison with the earlier ones, and finally, in Section~\ref{s:conclusion} we will summarize our findings.  

\section{Data}
\label{s:data}
\begin{figure}[h!]
\begin{center}
\includegraphics[width=\textwidth]{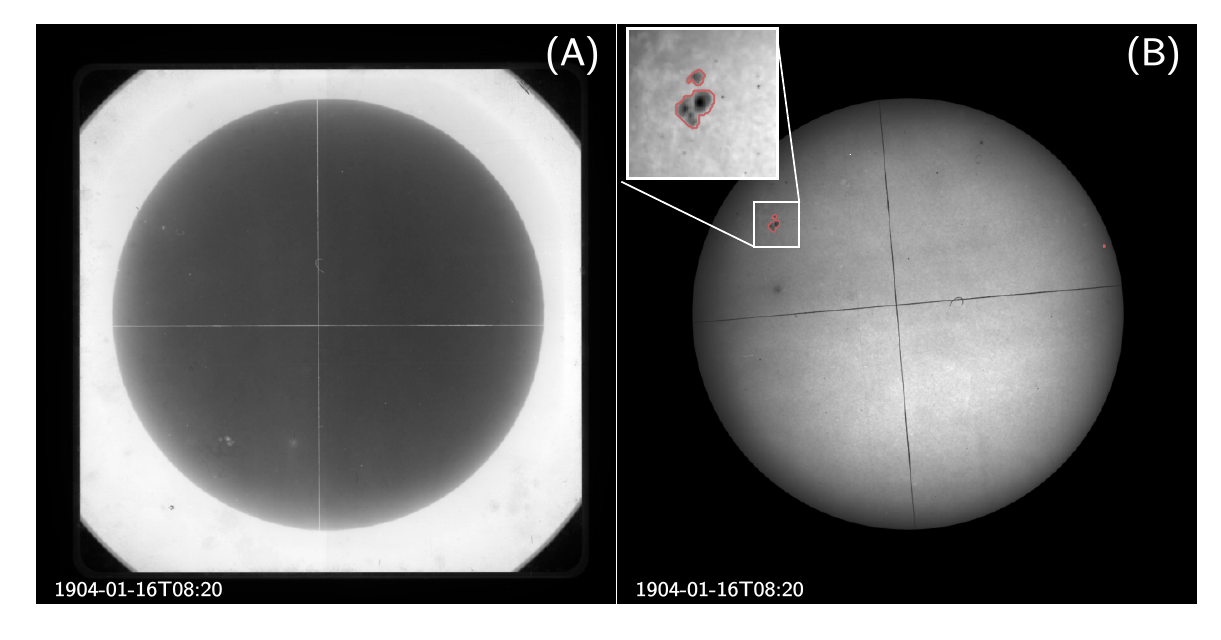}
\end{center}
\caption{A representative example of a very first observation at KoSO, dated 16th January 1904 (A) raw digitised negative image and (B) calibrated image with the contour of sunspot regions detected using a semi-automated method. A zoomed-in view of the detected sunspot is shown in the inset.}
\label{fig1:context_image}
\end{figure}

 The  white-light observation at Kodaikanal started in 1904 using a 10~cm objective lens telescope which was later replaced in 1912 by a better quality lens while maintaining same size. A few years later, a 15~cm achromatic lens was installed, and the same setup has been used to take observations since June 13th, 1918 \citep{Sivaraman1993}, and is still in use. Since 2017, the unavailability of photographic films has interrupted regular observation, but it is still taken whenever the films are available. These observations at KoSO, initially taken on the photographic plates/films, have been digitized using 4k$\times$4k CCD at Kodaikanal and made available for the community \citep{Ravindra2013, Mandal2017a}. Here, we use the white-light digitized data for the period of 1904\,--\,1920 and 2012\,--\,2018, and will extend the sunspot-area series reported earlier \citep[1921\,--\,2011;][]{Mandal2017a} for the period of 1904\,--\,2017 (114~years , covering $\approx$ 11 solar cycles). An example of a digitized white-light image from the very initial period, is shown in Figure~\ref{fig1:context_image}A. Furthermore, we discovered that there are a few observations during 1921\,--\,2011, that were missed in earlier area series \citep{Mandal2017a}, so we included them in the updated and revised series.

 \subsection{Updated Data Statistics}
 
 In Figure~\ref{fig2:data_statistics}(A), we present the number of observations per year from 1904\,--\,2017, with grey histograms representing the observations used in area-series as reported by \citet{Mandal2017a} and blue ones representing the additional observations included in this new series. Since, Figure~\ref{fig2:data_statistics}(A) includes the multiple observation from the same day hence, to get an idea about the data coverage, we counted the number of observing days in each year and plotted it against the years in Figure~\ref{fig2:data_statistics}(B). From Figure~\ref{fig2:data_statistics}(B), we infer that the KoSO provides continuous white-light observations in the last 114~years with coverage of $\approx$73\%. However, Kodaikanal experiences a rainy season each year during July to November and this has an impact on the data coverage \citep[see ][]{Bappu1967}.

\subsubsection{{\it Where does the KoSO stand?}}
\begin{figure}[h!]
\begin{center}
\includegraphics[width=\textwidth]{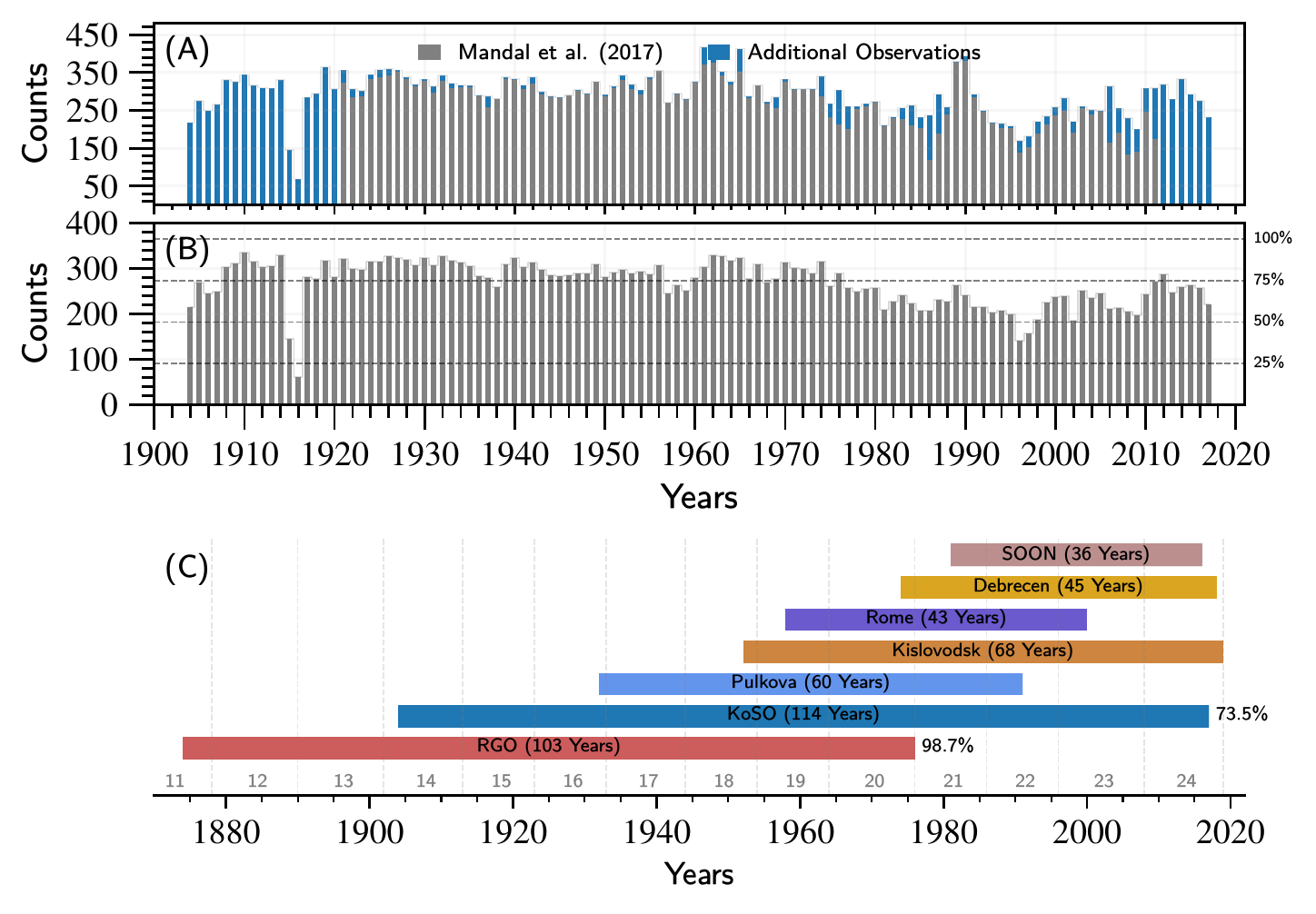}
\end{center}
\caption{Data statistics showing (A) the number of observations per year for each year where two colours (grey) data used in \citet{Mandal2017a} and (blue) the additional observation that has been added to the series; (B) the number of observing days in a year after including all the data for 1904\,--\,2017. (C) is showing the data coverage for different observatories, including KoSO and RGO.}
\label{fig2:data_statistics}
\end{figure}

In Figure~\ref{fig2:data_statistics}C, we compare the extent of data from the various observatories, showing that the KoSO has one of the most extended white-light data series (114~years) in the world. After KoSO, RGO provides the longest white-light sunspot data series (103~years). Although KoSO offers the most extended series, if we look at the percentage data coverage, RGO stands far ahead of the KoSO with data coverage of $\approx 98.7\%$ compared to $73.5\%$ of KoSO. Despite the fact that RGO has better coverage of the data but it has compiled the data from various observatories around the globe including KoSO \citep{Willis2013}, where as KoSO provides the white-light series for 114~years observed from same location as well as same setup (since 1918). Hence, the consistency and homogeneity in KoSO white-light sunspot data make it a unique and well-suited resource for the long-term study of the Sun.

\subsection{Data Calibration}

We follow the same calibration steps, which include flat fielding, disk detection and disk centring--to bring the disk centre to the image centre, as described in \citet{Ravindra2013}. Here we also like to point out that in the year 1920, we notice something very peculiar, the size of the disk is larger by more than 400 pixels compared to the rest of the images in the archive (see Figure~S1 (A) and (B) in the supplementary material for example). This difference is because, during digitization, the camera has been moved a bit closer to the plate \citep[see][for the details of digitizer unit]{Ravindra2013} leading to the bigger size of the disk for this particular year. 

Now, the next step is to correctly orient these images to get the solar North at the top of the image. In the earlier work by \citet{Ravindra2013} and \citet{Mandal2017a}, this correction has been performed using the East-West line present in the image as a reference line. The same technique is used here to correctly rotate the digitized images from the period of 2012\,--\,2017. However, this technique can not be used for the data from 1904\ to 1920 as they suffer from two different inconsistencies. Firstly, during 1904\,--\,1908 (till 11 September 1908), images have two perpendicular straight lines, as seen in Figure~\ref{fig1:context_image}(A), which represent the geographical East-West (EW) and the North-South (NS) direction, but the problem is we do not know which of them represents EW or NS; and secondly, during digitization, corresponding to the period of 1904\,--\,1920 accidentally images were flipped and rotated randomly, which makes the task of de-rotation even more challenging. Before we discuss the details of the alternative method that we used to get the proper orientation (discussed in Section~\ref{ss:orient}) of these effected images we go for the sunspot detection first. For that, we follow the same semi-automatic sunspot detection algorithm described in \citet{Ravindra2013} and \citet{Mandal2017a} and stored the information of sunspot regions in the form of binary masks. A representative example is shown in Figure~\ref{fig1:context_image}(B), where detected sunspot regions are marked using red contours.  

\subsubsection{{\it Orientation Correction}}
\label{ss:orient}
\begin{figure}[h!]
\begin{center}
\includegraphics[width=\textwidth]{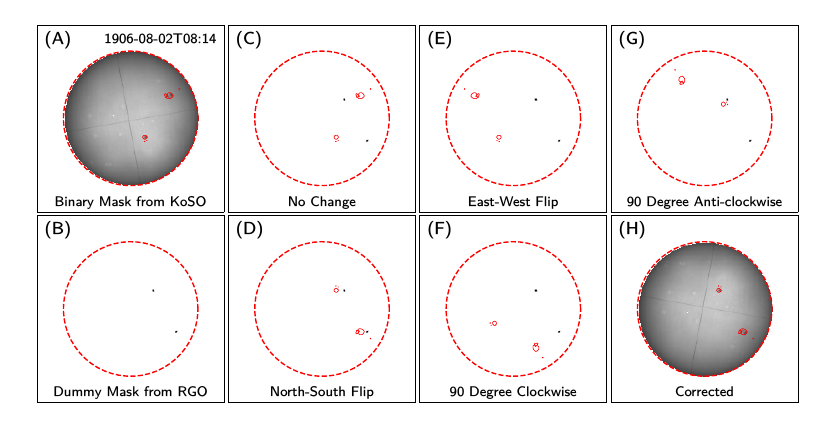}
\end{center}
\caption{A representative example shows the steps we use to identify the correct orientation. (A) shows the image we have taken from KoSO; here, red contours denote the identified sunspot regions. (B) shows the dummy mask we create from the near-simultaneous RGO sunspot data. Here, small dark regions represent the location of sunspots as per RGO observation. (C), (D), (E), (F) and (G) represents the binary mask of KoSO overlapped with RGO (dummy) mask for all five possible orientations. (H) shows the corrected digitized image; in this case, it is N-S flipped.}\label{fig:s1}
\end{figure}

So far, we have discussed the issue of image orientation, and now we will look at how we can get the correct orientation of the images for the aforementioned period. Here, we cannot use EW or NS reference line for the first five years of data (1904\,--\,1908) since we do not know which of them represents EW or NS. In principle, only the EW line in the rest of the data can be used as a reference to get proper orientation. However, these images were flipped in the EW or NS direction during the digitization process, so we have to look for an alternate method. For this we used the already available sunspot location information from the RGO  digitized full-disk images. In this method, firstly we chose the closest observation from the RGO data series, noted the time difference between observations is less than 12~hours. Then using the available sunspot heliographic location information, we create a dummy mask with the same image size as the KoSO sunspot detected binary mask. After that, the sunspot binary mask obtained from the KoSO data is overlapped with the dummy mask created from RGO data for five different possibilities (i) no change, i.e. correct orientation, (ii) North-South flip, (iii) East-West flip, (iv) \ang{90} clockwise and, (v) \ang{90} anti-clockwise. We also looked for the overlap and mark the observations with the appropriate flag. There are few cases where no overlap is seen in any of the five mentioned cases; hence we flag these observations as ``others.'' An example of the steps mentioned above is represented in Figure~\ref{fig:s1}(A) to {fig:s1}(H).

Now, following the aforementioned steps, we go through 3565 observations during 1904\,--\,1920 and flag them accordingly, as represented in Table~\ref{table1}. We see in Table~\ref{table1} that the majority ($\approx53\%$, predominantly in 1904\,--\,1912) of the observation are NS flipped (for yearly distribution of different orientations, see Figure~S3 in supplementary material). In Figure~\ref{fig3:rotation_example} (A) and \ref{fig3:rotation_example}(B), we show two such cases where the images are EW and NS flipped, respectively. Apart from that, we also notice that there are 73  ($2\%$) observations for which we cannot get the correct orientation. When we carefully looked at them, we found that the most probable reason is the incorrect time of observation of these images, which leads to fallacious overlapping pair. The unavailability of the correct time of observation of these images makes it difficult to get an accurate orientation, and therefore, we do not include these observations in our analysis.

\begin{table}
\begin{center}
\caption{Number of observations for different flags representing the orientation of the images during 1904\,--\,1920.}
\label{table1}
\begin{tabular}{r c r}
\toprule
Orientation & Number of Observations & Percentage\\
\midrule
Total Number of Observations & 3565 &\\
Correct & 1576 &44.2\%\\
North-South Flip & 1892 &53.1\%\\
East-West Flip & 19 &$0.53$\%\\
\ang{90} (Clock-wise) & 3 & $<0.1$\%\\
\ang{-90} (Anti-clockwise)& 2 &$<0.1$\%\\
Others & 73 &2.0\%\\
\bottomrule
\end{tabular}
\end{center}
\end{table}
\begin{figure}[h!]
\begin{center}
\includegraphics[width=\textwidth]{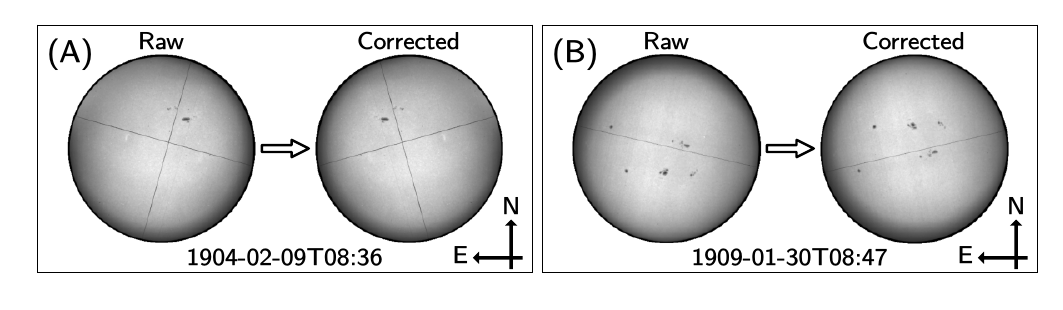}
\end{center}
\caption{Two representative examples from 1904 and 1909 show the orientation correction (A) for East-West flip and (B) North-South flip images.}
\label{fig3:rotation_example}
\end{figure}


\section{Results}
\label{s:result}

To detect the sunspots, we use a modified version of the Sunspot Tracking And Recognition Algorithm \citep[STARA, ][]{Watson2011}. Our algorithm is a semiautomated one and has previously been used successfully on KoSO white-light images by \citet{Ravindra2013} and \citet{Mandal2017a}. It is important to note here that although the observing setup at KoSO was modified several times between 1904\,--\,1918, it has no effect on our sunspot detection algorithm (see Figure~S3 in supplementary material). Lastly, for each detected spot, we calculate its area (corrected for projection) and heliographic coordinates (latitude and longitude).

First and foremost, we compare the sunspot area obtained from KoSO digitized white-light data with the existing sunspot area series. Then we discuss the sunspot umbra area extracted from them.

\subsection{Sunspot Area Series}
\subsubsection{{\it Comparison with Existing Series}}
\begin{figure}[h!]
\begin{center}
\includegraphics[width=\textwidth]{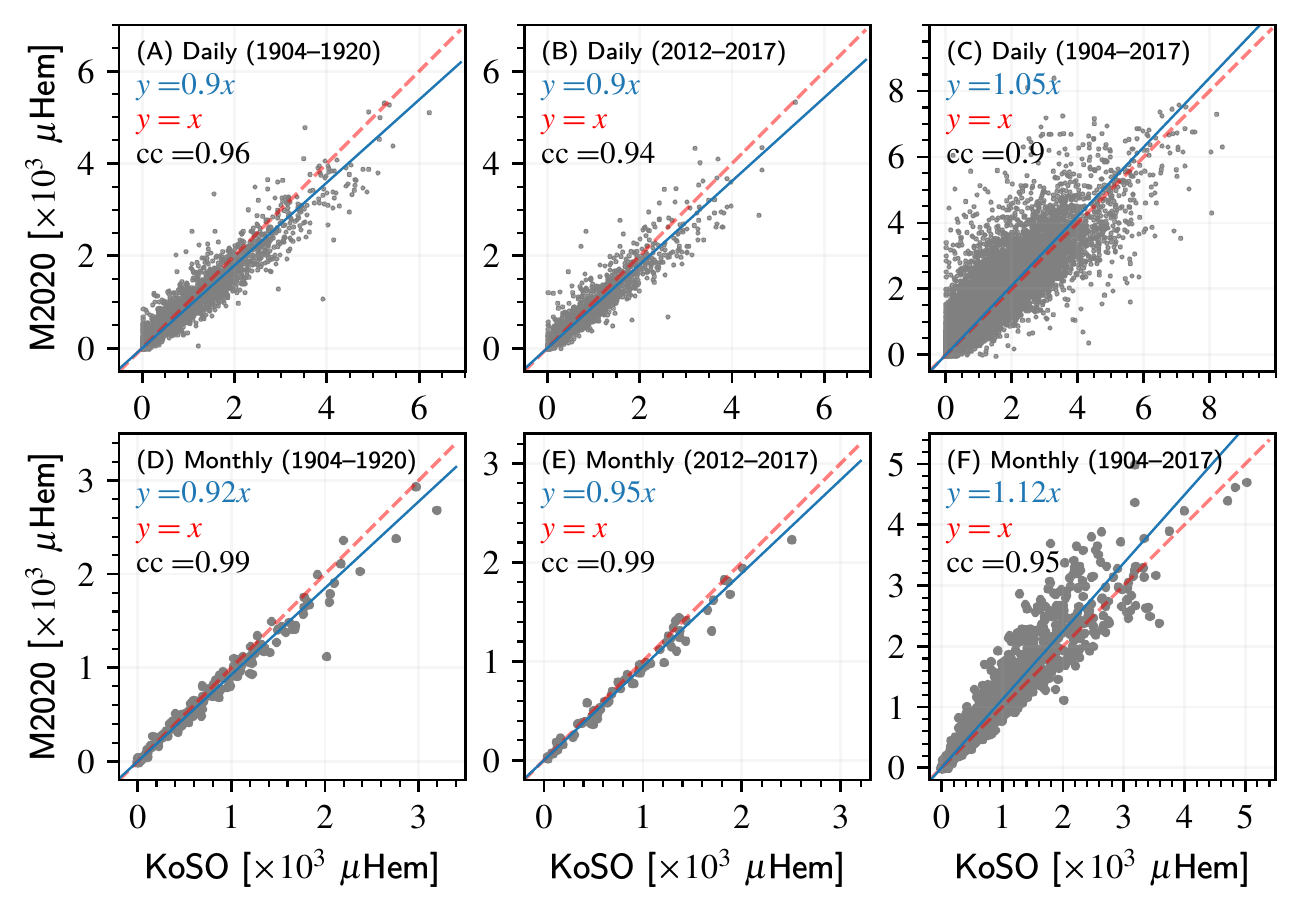}
\end{center}
\caption{The scatter plot showing the comparison of the daily sunspot area, panels (A), (B) and (C); and monthly averaged sunspot area, panels (D), (E) and (F) with M2020 for 1904\,--\,1920, 2012\,--\,2017 and for the extended period of 1904\,--\,2017. In each panel, the line with unit slop and the linear fit with zero intercepts are represented by solid red and blue dashed lines, respectively.}\label{fig4:area_comparision}
\end{figure}

In Figure~\ref{fig4:area_comparision}(A) and \ref{fig4:area_comparision}(B) we compare the daily sunspot area obtained from KoSO with the recently cross calibrated composite sunspot area series from \citet{Mandal2020} (M2020 hereafter) for 1904\,--\,1920 and 2012\,--\,2017, respectively. In M2020 the sunspot area data series for 1904\,--\,1920 primarily contains data from the RGO photographic results, whereas for 2012\,--\,2017 it is from Debrecen Photographic Data (DPD). The daily sunspot data show a good correlation with M2020 in both periods, with a correlation coefficient ($cc$) of more than 0.9. When we use linear fit ($y=mx$) for the daily sunspot data, the slope turns out to be 0.9 for both the periods 1904\,--\,1920 and 2012\,--\,2017. Slope less than unity signifies that, in KoSO data, we overestimated the area in these periods, whereas the earlier series \citep{Mandal2017a} and recently M2020 reported underestimation for 1921\,--\,2011. In Figure~\ref{fig4:area_comparision}(C), we compare the KoSO daily sunspot area data for the whole period (1904\,--\,2017) with M2020, and we see the underestimation of the sunspot area in KoSO as it is dominated by 80\% of data (1921\,--\,2011) where it is undervalued. In Figures~\ref{fig4:area_comparision}(D),  \ref{fig4:area_comparision}(E) and \ref{fig4:area_comparision}(F), we see similar properties with monthly averaged data but with slightly higher $cc$ values, which is obvious because averaging will reduce the scatter. The possible reasons for this difference could be (i) the data quality, which is not uniform through the whole data span, (ii) degradation of plates/films with time, and (iii) presence of artefacts, e.g., scratch, dust etc. In addition to the availability of data for Cycle~14 and Cycle~15, the outstanding homogeneity of the KoSO white-light data makes this data an excellent resource to cross-correlate RGO sunspot data with KoSO sunspot area data in overlapping periods.

\subsubsection{{\it Extended Sunspot Area Series}}

\begin{figure}[h!]
\begin{center}
\includegraphics[width=\textwidth]{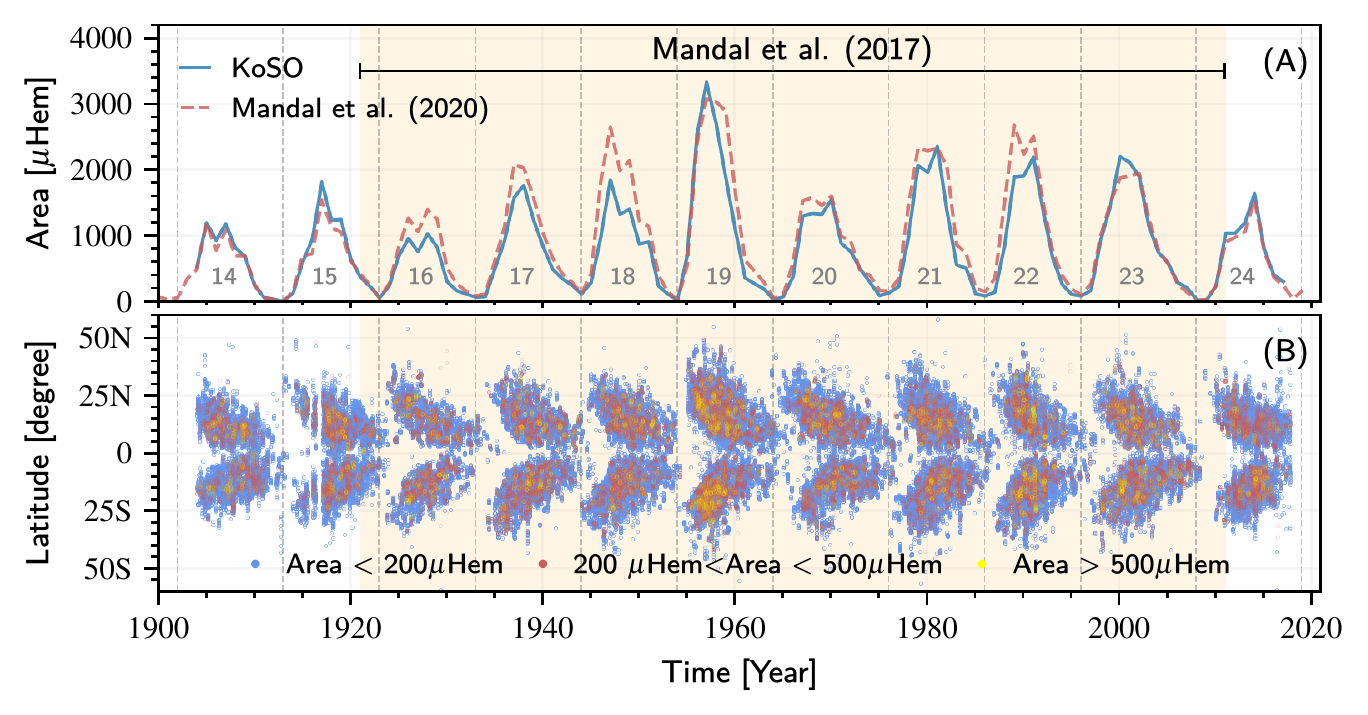}
\end{center}
\caption{ In panel (A) yearly averaged sunspot area obtained from KoSO white-light digitized data (solid blue) is plotted as a function of time along with the yearly averaged M2020 (dashed red). Panel (B) show the latitude time plot for the extended period obtained from KoSO sunspot data.}\label{fig5:area_series}
\end{figure}

Now, in Figure~\ref{fig5:area_series}(A), we show the variation of yearly averaged sunspot area from KoSO for the extended period of 1904\,--\,2017 (covering 11 solar cycles, Cycle~14 to Cycle~24) along with M2020. It is evident from Figure~\ref{fig5:area_series}(A) that the new addition in the series shows a good match with M2020 during 1904\,--\,1920 and 2012\,--\,2017 whereas in 1921\,--\,2011, the yearly averaged value is lower than M2020, and the possible reasons have been already been reported and discussed in \citet{Mandal2017a}. In Figure~\ref{fig5:area_series}(B), we also show the latitude-time plot, the so-called butterfly diagram, for the extended period of 1904\,--\,2017. The important point to notice here is the butterfly diagram for 1904\,--\,1920 giving us the correct representation of latitude-time plot consistent with other sunspot data series, M2020. This was the period in which images were corrected for random flip and rotation; therefore, the consistency of the butterfly diagram verifies orientation corrections applied. This extended and updated area series, along with orientation information for the initial 17 years, will be publicly available for community use. 

\subsection{Umbra Area Series}

\begin{figure}[h!]
\begin{center}
\includegraphics[width=\textwidth]{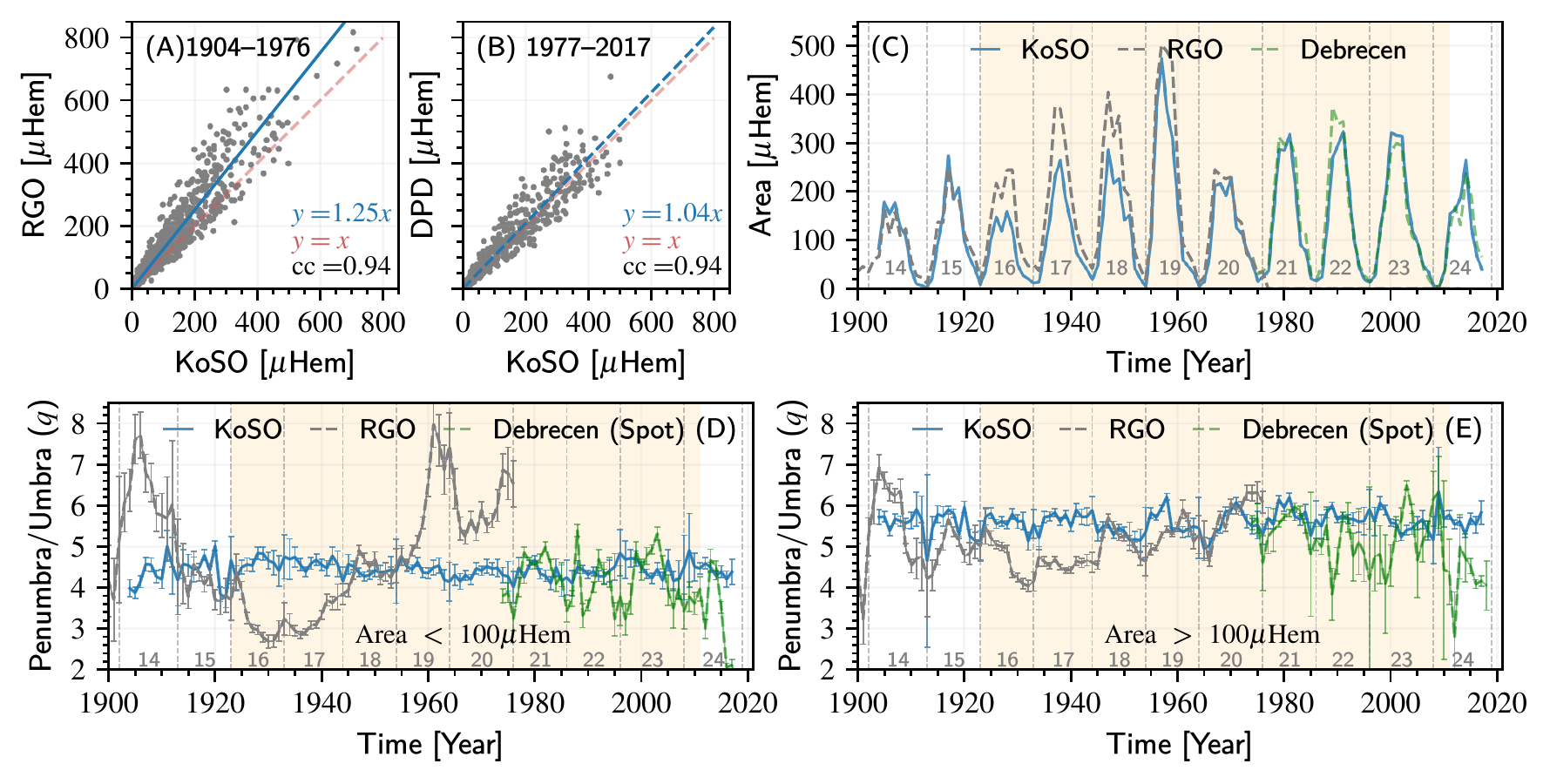}
\end{center}
\caption{In panels (A) and (B), we compare the KoSO monthly averaged umbra area with RGO and DPD umbra, respectively, in their corresponding periods. Panel (C) shows the solar cycle variation of yearly averaged umbra area compared to RGO and DPD umbra area. Panel (D) and (E) show the variation of $q$ for smaller and larger sunspots for an extended period, respectively. The shaded regions represent the data used in \citet{Jha2019}.}\label{fig6:umbra_penumbra}
\end{figure}

Following the automatic umbra detection method as explained in \citet{Jha2019}, we detected the umbra area from the updated and extended white-light data series. In Figure~\ref{fig6:umbra_penumbra}(A), we show the scatter plot between the monthly averaged umbra area from KoSO and RGO for 1904\,--\,1976. After 1976 RGO does not provide sunspot (and umbral) area data (as observation program has been transferred to Debrecen Observatory), hence we use the DPD umbral area in 1977\,--\,2017 to compare with KoSO, as shown in Figure~\ref{fig6:umbra_penumbra}(B). Both the data (RGO \& DPD) show a very good correlation (0.94 \& 0.95) with the KoSO umbra area in their corresponding periods. Despite the fact that the correlation coefficient is quite good, we can infer from the fitted line that there is a considerable underestimation of the umbra area in KoSO compared to RGO, which is not so severe in the case of DPD. When we compare the yearly average by plotting all of them simultaneously in Figure~\ref{fig6:umbra_penumbra}(C), we notice that this underestimation is primarily coming from 3 cycles (Cycle~16, Cycle~17 and Cycle~18). For the same set of cycles, we also see a bit of undervalued sunspot area in Figure~\ref{fig5:area_series}(A). 

\subsubsection{{\it Penumbra to Umbra Area Ratio}}
In this subsection, we extend the work presented in \citet{Jha2019} by calculating the penumbra to umbra area ratio ($q$) for the extended period. 
In Figures~\ref{fig6:umbra_penumbra}(D) and \ref{fig6:umbra_penumbra}(E), we show the variation in $q$ with time for smaller (Area $<100~\mu$Hem) and larger (Area $>100~\mu$hem) sunspots for the complete data span. Furthermore, in Figure~\ref{fig6:umbra_penumbra}(D) and \ref{fig6:umbra_penumbra}(E), we do not find any abrupt change in $q$ between 1904 and 1918 (it was the period when the telescope had been modified several times), which again demonstrate the robustness of sunspot detection algorithm (and subsequently umbra detection technique).

In Figure~\ref{fig6:umbra_penumbra}(D), we do not notice any variation in $q$ for smaller spots even in the extended period, which is consistent with the findings of \citet{Jha2019}. Furthermore, the variation of $q$ for larger sunspots is also consistent with the earlier findings of \citet{Jha2019} as well as \citet{Hathaway2013}. We do note, however, that in KoSO, the spots are yet to be classified into groups, unlike RGO. Hence, there may be a slight variation in the absolute value of $q$ if we consider individual spots. To check this hypothesis, we over-plot the $q$ values calculated from DPD, which provides information of individual spots. And indeed, we find that although the absolute value of $q$ is slightly less than that of KoSO, the trend remains similar in both the data sets. Furthermore,  \citet{Carrasco2018}, and recently \citet{Hou2022} (including all spots altogether) did not find any such trend in $q$ as was seen in RGO data.

\section{Summary and Conclusion}
\label{s:conclusion}
In this article, we present the sunspot area series from KoSO, for the period of 1904\,--\,1920, and 2012\,--\,2017, which were not included due to calibration issues and incomplete digitization in the earlier published series \citep{Mandal2017a}. The inclusion of these data has extended the sunspot area series reported in \citet{Mandal2017a} for 1904\,--\,2017 (114~years, covering $\approx11$ solar cycles). We first resolved the calibration issues with the initial period (1904\,--\,1920) of data, primarily the issue with orientation and $\approx 53\%$ of images were NS flipped. After orientation correction, we detected the sunspot using a semi-automated method and identified the umbral regions using a completely automatic algorithm. By using these automatic or semi-automatic algorithms, we ensure minimal or no human subjectivity in these detection processes. After that, we compared the daily and monthly averaged sunspot area with M2020 \citep{Mandal2020}, which show a good correlation with $cc>0.9$ in both these period (1904\,-- ,1920 and 2012\,--\,2017). The solar cycle variation of yearly averaged sunspot area shows excellent agreement over the extended period of 1904\,--\,2017. Since RGO has collected data from different observatories to fill the gaps, it provides significantly higher data coverage than KoSO, which has acquired all the data from the same location. It makes KoSO one of the most homogeneous and unique data series for such an extended period.

We have also compared the yearly averaged umbra area series with RGO (1904\,--\,1976) and DPD (1977\,--\,2017) in their corresponding periods, which show a good agreement between the umbral areas (cc$>0.9$) except in 3 cycles (Cycle~15, Cycle~16 and Cycle~17) where we notice KoSO umbral area significantly lower than RGO umbral area. We also calculated the penumbra to umbra area ratio ($q$) and compared it with RGO data. We noted that KoSO data do not show any long-term trend in the ratio for smaller (Area $<100~\mu$hem) as well as for larger (Area $>100~\mu$hem), which further supports the findings of \citet{Jha2019}. In addition, we also note that the change of observing setup during 1904\,--\,1918 has not affected our ability to detect sunspots or umbra and hence our results.

The availability of high-resolution white-light digitized data for Cycle~14 and Cycle~15 is a key asset for the long-term studies of the Sun, and it provides an excellent opportunity to cross-correlate the sunspot area data for observatories. In addition, the accessibility of a homogeneous and uniform white-light digitized data for such a long period observed from the same location and setup will benefit the community for the long-term studies of the Sun and its global magnetic field variability. Moreover, KoSO also provides sun charts, which combine the multi-wavelength observation in a single drawing. These sun charts are getting digitized and will help fill the gaps and make the series even more homogeneous. In future, we will be looking for machine learning (ML), artificial intelligence (AI) and deep learning based methods to further improve the sunspot detection method by making it completely automatic. Furthermore, ML and AI based method can be expended to the study of historical global solar magnetic field and space weather conditions.

The digitized white-light data and the area series for the extended period will be available at \url{https://kso.iiap.res.in/new/data}. The sunspot area series for the extended period presented here is also available at \url{https://github.com/bibhuraushan/KoSoDigitalArchive}.

\section*{Conflict of Interest Statement}

The authors declare that the research was conducted in the absence of any commercial or financial relationships that could be construed as a potential conflict of interest.

\section*{Author Contributions}
BKJ has done the data calibration (orientation correction), umbra detection using an automatic algorithm, all the analysis presented and manuscript drafting. MH has done 80\% of sunspot detection, and AP has done the rest 20\% of sunspot detection and contributed to the calibration step of orientation correction. BR has performed the data's basic calibration (flat fielding and disk detection) step. SM and DB have helped in the structuring of the draft and helped in the language correction.

\section*{Funding}
 This project is partially funded by the project grant DST/ICPS/CLUSTER/DataScience/2018/General/Sl. No.18.

\section*{Acknowledgments}
Kodaikanal Solar Observatory is a facility of the Indian Institute of Astrophysics, Bangalore, India. This data is now available for public use at http://kso.iiap.res.in through a service developed at IUCAA under the Data Driven Initiatives project funded by the National Knowledge Network. BKJ would like to thank the entire digitization team for their tireless effort to digitize the historical data and make data publically available.


\section*{Data Availability Statement}
The dataset ANALYZED and Generated for this study can be found in the \url{https://kso.iiap.res.in/new/data}. The generated data can be also accessed via \url{https://github.com/bibhuraushan/KoSoDigitalArchive}. Further enquiry about the data can be directed to \href{mailto:dipu@iiap.res.in}{dipu@iiap.res.in}.

\bibliographystyle{Frontiers-Harvard} 

\begin{thebibliography}{30}
\providecommand{\natexlab}[1]{#1}
\expandafter\ifx\csname urlstyle\endcsname\relax
  \providecommand{\doi}[1]{doi:\discretionary{}{}{}#1}\else
  \providecommand{\doi}{doi:\discretionary{}{}{}\begingroup
  \urlstyle{rm}\Url}\fi
\providecommand{\selectlanguage}[1]{\relax}
\providecommand{\bibAnnoteFile}[1]{%
  \IfFileExists{#1}{\begin{quotation}\noindent\textsc{Key:} #1\\
  \textsc{Annotation:}\ \input{#1}\end{quotation}}{}}
\providecommand{\bibAnnote}[2]{%
  \begin{quotation}\noindent\textsc{Key:} #1\\
  \textsc{Annotation:}\ #2\end{quotation}}

\bibitem[{{Balmaceda} et~al.(2009){Balmaceda}, {Solanki}, {Krivova}, and
  {Foster}}]{Balmaceda2009}
{Balmaceda}, L.~A., {Solanki}, S.~K., {Krivova}, N.~A., and {Foster}, S.
  (2009).
\newblock {A homogeneous database of sunspot areas covering more than 130
  years}.
\newblock \emph{Journal of Geophysical Research (Space Physics)} 114, A07104.
\newblock \doi{10.1029/2009JA014299}
\input{Balmaceda2009}

\bibitem[{{Bappu}(1967)}]{Bappu1967}
{Bappu}, M.~K.~V. (1967).
\newblock {Solar Physics at Kodaikanal}.
\newblock \emph{\solphys} 1, 151--156.
\newblock \doi{10.1007/BF00150312}
\input{Bappu1967}

\bibitem[{{Baranyi} et~al.(2001){Baranyi}, {Gyori}, {Ludm{\'a}ny}, and
  {Coffey}}]{Baranyi2001}
{Baranyi}, T., {Gyori}, L., {Ludm{\'a}ny}, A., and {Coffey}, H.~E. (2001).
\newblock {Comparison of sunspot area data bases}.
\newblock \emph{\mnras} 323, 223--230.
\newblock \doi{10.1046/j.1365-8711.2001.04195.x}
\input{Baranyi2001}

\bibitem[{{Baranyi} et~al.(2013){Baranyi}, {Kir{\'a}ly}, and
  {Coffey}}]{Baranyi2013}
{Baranyi}, T., {Kir{\'a}ly}, S., and {Coffey}, H.~E. (2013).
\newblock {Indirect comparison of Debrecen and Greenwich daily sums of sunspot
  areas}.
\newblock \emph{\mnras} 434, 1713--1720.
\newblock \doi{10.1093/mnras/stt1134}
\input{Baranyi2013}

\bibitem[{{Carrasco} et~al.(2018){Carrasco}, {Vaquero}, {Trigo}, and
  {Gallego}}]{Carrasco2018}
{Carrasco}, V.~M.~S., {Vaquero}, J.~M., {Trigo}, R.~M., and {Gallego}, M.~C.
  (2018).
\newblock {A Curious History of Sunspot Penumbrae: An Update}.
\newblock \emph{\solphys} 293, 104.
\newblock \doi{10.1007/s11207-018-1328-z}
\input{Carrasco2018}

\bibitem[{{Charbonneau}(2010)}]{Charbonneau2010}
{Charbonneau}, P. (2010).
\newblock {Dynamo Models of the Solar Cycle}.
\newblock \emph{Liv. Rev. Sol. Phys.} 7, 3.
\newblock \doi{10.12942/lrsp-2010-3}
\input{Charbonneau2010}

\bibitem[{{Fligge} and {Solanki}(1997)}]{Fligge1997}
{Fligge}, M. and {Solanki}, S.~K. (1997).
\newblock {Inter-Cycle Variations of Solar Irradiance: Sunspot Areas as a
  Pointer}.
\newblock \emph{\solphys} 173, 427--439.
\newblock \doi{10.1023/A:1004971807172}
\input{Fligge1997}

\bibitem[{{Hale}(1908)}]{Hale1908}
{Hale}, G.~E. (1908).
\newblock {On the Probable Existence of a Magnetic Field in Sun-Spots}.
\newblock \emph{\apj} 28, 315.
\newblock \doi{10.1086/141602}
\input{Hale1908}

\bibitem[{{Hasan} et~al.(2010){Hasan}, {Mallik}, {Bagare}, and
  {Rajaguru}}]{Hasan2010}
{Hasan}, S.~S., {Mallik}, D.~C.~V., {Bagare}, S.~P., and {Rajaguru}, S.~P.
  (2010).
\newblock {Solar Physics at the Kodaikanal Observatory: A Historical
  Perspective}.
\newblock In \emph{Magnetic Coupling between the Interior and Atmosphere of the
  Sun}. vol.~19 of \emph{Astrophysics and Space Science Proceedings}, 12--36.
\newblock \doi{10.1007/978-3-642-02859-5\_3}
\input{Hasan2010}

\bibitem[{{Hathaway}(2013)}]{Hathaway2013}
{Hathaway}, D.~H. (2013).
\newblock {A Curious History of Sunspot Penumbrae}.
\newblock \emph{\solphys} 286, 347--356.
\newblock \doi{10.1007/s11207-013-0291-y}
\input{Hathaway2013}

\bibitem[{Hathaway(2015)}]{Hathaway2015}
Hathaway, D.~H. (2015).
\newblock The solar cycle.
\newblock \emph{\lrsp} 12, 4.
\newblock \doi{10.1007/lrsp-2015-4}
\input{Hathaway2015}

\bibitem[{{Hou} et~al.(2022){Hou}, {Zeng}, {Zheng}, {Luo}, {Deng}, {Li}
  et~al.}]{Hou2022}
{Hou}, J.-W., {Zeng}, S.-G., {Zheng}, S., {Luo}, X.-Y., {Deng}, L.-H., {Li},
  Y.-Y., et~al. (2022).
\newblock {Chinese Sunspot Drawings and Their Digitization-(VII) Sunspot
  Penumbra to Umbra Area Ratio Using the Hand-Drawing Records from Yunnan
  Observatories}.
\newblock \emph{Research in Astronomy and Astrophysics} 22, 095012.
\newblock \doi{10.1088/1674-4527/ac7f87}
\input{Hou2022}

\bibitem[{Jha et~al.(2018)Jha, Mandal, and Banerjee}]{Jha2018}
Jha, B.~K., Mandal, S., and Banerjee, D. (2018).
\newblock Long-term variation of sunspot penumbra to umbra area ratio: A study
  using kodaikanal white-light digitized data.
\newblock \emph{Proceedings of the International Astronomical Union} 13,
  185–186.
\newblock \doi{10.1017/S1743921318001989}
\input{Jha2018}

\bibitem[{{Jha} et~al.(2019){Jha}, {Mandal}, and {Banerjee}}]{Jha2019}
{Jha}, B.~K., {Mandal}, S., and {Banerjee}, D. (2019).
\newblock {Study of Sunspot Penumbra to Umbra Area Ratio Using Kodaikanal
  White-light Digitised Data}.
\newblock \emph{\solphys} 294, 72.
\newblock \doi{10.1007/s11207-019-1462-2}
\input{Jha2019}

\bibitem[{{Jha} et~al.(2021){Jha}, {Priyadarshi}, {Mandal}, {Chatterjee}, and
  {Banerjee}}]{Jha2021}
{Jha}, B.~K., {Priyadarshi}, A., {Mandal}, S., {Chatterjee}, S., and
  {Banerjee}, D. (2021).
\newblock {Measurements of Solar Differential Rotation Using the Century Long
  Kodaikanal Sunspot Data}.
\newblock \emph{\solphys} 296, 25.
\newblock \doi{10.1007/s11207-021-01767-8}
\input{Jha2021}

\bibitem[{{Jiang} et~al.(2011){Jiang}, {Cameron}, {Schmitt}, and
  {Sch{\"u}ssler}}]{Jiang2011}
{Jiang}, J., {Cameron}, R.~H., {Schmitt}, D., and {Sch{\"u}ssler}, M. (2011).
\newblock {The solar magnetic field since 1700. II. Physical reconstruction of
  total, polar and open flux}.
\newblock \emph{\aap} 528, A83.
\newblock \doi{10.1051/0004-6361/201016168}
\input{Jiang2011}

\bibitem[{{Jiang} et~al.(2014){Jiang}, {Hathaway}, {Cameron}, {Solanki},
  {Gizon}, and {Upton}}]{Jiang2014a}
{Jiang}, J., {Hathaway}, D.~H., {Cameron}, R.~H., {Solanki}, S.~K., {Gizon},
  L., and {Upton}, L. (2014).
\newblock {Magnetic Flux Transport at the Solar Surface}.
\newblock \emph{\ssr} 186, 491--523.
\newblock \doi{10.1007/s11214-014-0083-1}
\input{Jiang2014a}

\bibitem[{{Mandal} et~al.(2017){Mandal}, {Hegde}, {Samanta}, {Hazra},
  {Banerjee}, and {Ravindra}}]{Mandal2017a}
{Mandal}, S., {Hegde}, M., {Samanta}, T., {Hazra}, G., {Banerjee}, D., and
  {Ravindra}, B. (2017).
\newblock {Kodaikanal digitized white-light data archive (1921-2011): Analysis
  of various solar cycle features}.
\newblock \emph{\aap} 601, A106.
\newblock \doi{10.1051/0004-6361/201628651}
\input{Mandal2017a}

\bibitem[{{Mandal} et~al.(2020){Mandal}, {Krivova}, {Solanki}, {Sinha}, and
  {Banerjee}}]{Mandal2020}
{Mandal}, S., {Krivova}, N.~A., {Solanki}, S.~K., {Sinha}, N., and {Banerjee},
  D. (2020).
\newblock {Sunspot area catalog revisited: Daily cross-calibrated areas since
  1874}.
\newblock \emph{\aap} 640, A78.
\newblock \doi{10.1051/0004-6361/202037547}
\input{Mandal2020}

\bibitem[{{Nagovitsyn} et~al.(2017){Nagovitsyn}, {Pevtsov}, and
  {Osipova}}]{Nagovitsyn2017}
{Nagovitsyn}, Y.~A., {Pevtsov}, A.~A., and {Osipova}, A.~A. (2017).
\newblock {Long-term variations in sunspot magnetic field-area relation}.
\newblock \emph{Astronomische Nachrichten} 338, 26--34.
\newblock \doi{10.1002/asna.201613035}
\input{Nagovitsyn2017}

\bibitem[{{Parker}(1955{\natexlab{a}})}]{Parker1955}
{Parker}, E.~N. (1955{\natexlab{a}}).
\newblock {Hydromagnetic Dynamo Models.}
\newblock \emph{\apj} 122, 293.
\newblock \doi{10.1086/146087}
\input{Parker1955}

\bibitem[{{Parker}(1955{\natexlab{b}})}]{Parker1955a}
{Parker}, E.~N. (1955{\natexlab{b}}).
\newblock {The Formation of Sunspots from the Solar Toroidal Field.}
\newblock \emph{\apj} 121, 491.
\newblock \doi{10.1086/146010}
\input{Parker1955a}

\bibitem[{{Parker}(1975)}]{Parker1975}
{Parker}, E.~N. (1975).
\newblock {The generation of magnetic fields in astrophysical bodies. X -
  Magnetic buoyancy and the solar dynamo}.
\newblock \emph{\apj} 198, 205--209.
\newblock \doi{10.1086/153593}
\input{Parker1975}

\bibitem[{{Ravindra} et~al.(2013){Ravindra}, {Priya}, {Amareswari}, {Priyal},
  {Nazia}, and {Banerjee}}]{Ravindra2013}
{Ravindra}, B., {Priya}, T.~G., {Amareswari}, K., {Priyal}, M., {Nazia}, A.~A.,
  and {Banerjee}, D. (2013).
\newblock {Digitized archive of the Kodaikanal images: Representative results
  of solar cycle variation from sunspot area determination}.
\newblock \emph{\aap} 550, A19.
\newblock \doi{10.1051/0004-6361/201220416}
\input{Ravindra2013}

\bibitem[{{Schwabe}(1844)}]{Schwabe1844}
{Schwabe}, H. (1844).
\newblock {Sonnenbeobachtungen im Jahre 1843. Von Herrn Hofrath Schwabe in
  Dessau}.
\newblock \emph{Astronomische Nachrichten} 21, 233.
\newblock \doi{10.1002/asna.18440211505}
\input{Schwabe1844}

\bibitem[{{Sivaraman} et~al.(1993){Sivaraman}, {Gupta}, and
  {Howard}}]{Sivaraman1993}
{Sivaraman}, K.~R., {Gupta}, S.~S., and {Howard}, R.~F. (1993).
\newblock {Measurement of Kodiakanal White-Light Images - Part One}.
\newblock \emph{\solphys} 146, 27--47.
\newblock \doi{10.1007/BF00662168}
\input{Sivaraman1993}

\bibitem[{Solanki(2003)}]{Solanki2003}
Solanki, S.~K. (2003).
\newblock Sunspots: An overview.
\newblock \emph{\aapr} 11, 153--286.
\newblock \doi{10.1007/s00159-003-0018-4}
\input{Solanki2003}

\bibitem[{{Tlatov} and {Pevtsov}(2014)}]{Tlatov2014}
{Tlatov}, A.~G. and {Pevtsov}, A.~A. (2014).
\newblock {Bimodal Distribution of Magnetic Fields and Areas of Sunspots}.
\newblock \emph{\solphys} 289, 1143--1152.
\newblock \doi{10.1007/s11207-013-0382-9}
\input{Tlatov2014}

\bibitem[{{Watson} et~al.(2011){Watson}, {Fletcher}, and
  {Marshall}}]{Watson2011}
{Watson}, F.~T., {Fletcher}, L., and {Marshall}, S. (2011).
\newblock {Evolution of sunspot properties during solar cycle 23}.
\newblock \emph{\aap} 533, A14.
\newblock \doi{10.1051/0004-6361/201116655}
\input{Watson2011}

\bibitem[{{Willis} et~al.(2013){Willis}, {Coffey}, {Henwood}, {Erwin}, {Hoyt},
  {Wild} et~al.}]{Willis2013}
{Willis}, D.~M., {Coffey}, H.~E., {Henwood}, R., {Erwin}, E.~H., {Hoyt}, D.~V.,
  {Wild}, M.~N., et~al. (2013).
\newblock {The Greenwich Photo-heliographic Results (1874 - 1976): Summary of
  the Observations, Applications, Datasets, Definitions and Errors}.
\newblock \emph{\solphys} 288, 117--139.
\newblock \doi{10.1007/s11207-013-0311-y}
\input{Willis2013}

\end{thebibliography}

\end{document}